\documentclass[12pt]{elsart}
\usepackage[english]{babel}
\usepackage{amsmath}
\usepackage{amssymb}

\begin{document}
\begin{frontmatter}
\title{Integrability conditions for nonautonomous quad-graph equations}

\author[India]{R. Sahadevan\thanksref{CAICYT}},
\author[UK]{O.G. Rasin} and
\author[UK]{P.E. Hydon}
\address[India]{Ramanujan Institute for Advanced Study in Mathematics,
 University of Madras, Chepauk, Chennai-600 005, Tamil Nadu, India}
\address[UK]{Department of Mathematics and Statistics, University of Surrey,
Guildford, Surrey GU2 7XH, UK}
\thanks[CAICYT]{Partially supported by the Indian National Science Academy and the Royal Society.}

\begin{abstract}
This paper presents a systematic investigation of the integrability
conditions for nonautonomous quad-graph maps, using the Lax pair
approach, the ultra-local singularity confinement criterion and
direct construction of conservation laws. We show that the
integrability conditions derived from each of the methods are the
one and the same, suggesting that there exists a deep connection
between these techniques for partial difference equations.\\

\end{abstract}

\end{frontmatter}
\section{Introduction}
Recently, there has been rapid development of research into
integrable discrete nonlinear systems governed by both ordinary
difference equations (mappings) and partial difference equations
(lattice equations) \cite{1,2,3,4,5}. It has been suggested that discrete
systems governed by difference equations are more fundamental than
the continuous ones described by differential equations. Efforts
have been made by several research groups to develop analytical
techniques to determine whether or not a given nonlinear partial
difference equation is integrable \cite{6,7,8,9}. It has been possible to
identify several rich classes of integrable nonlinear autonomous
partial difference equations that are amenable to solution by linear
methods \cite{9,10,11,12,13,14,SC0,RH0}. For a given autonomous integrable nonlinear partial
difference equation, the question of finding an integrable
nonautonomous version systematically is not yet solved. Examples of nonautonomous quad-graph equations are considered in \cite{SC0,GRS}; in particular \cite{GRS} deals with reductions to ordinary difference equations.

The aim of this article is to demonstrate that the
combination of the Lax pair formalism, the ultra-local singularity
confinement criterion and construction of conservation laws
provides an efficient tool to investigate integrable nonautonomous
nonlinear partial difference equations. We use these methods to find integrable nonautonomous
versions of the discrete Korteweg-de Vries (dKdV)
and modified discrete Korteweg-de Vries (dmKdV) equations.

The plan of the article is as follows. In \S 2 we
explain the Lax pair formalism for nonlinear partial difference
equations on the quad-graph and derive conditions for the existence of a Lax pair
for  nonautonomous dKdV and dmKdV equations. In \S 3 we derive conditions for each of
the above nonautonomous lattice equations under which the
ultra-local singularity confinement criterion is satisfied. Section 4
deals with the derivation of conditions on the above nonautonomous
lattice equations for which three-point conservation laws exist. In \S 5 we draw conclusions and
describe a possible direction of research.

\section{Lax pair formalism for partial difference equations}
In the following, $u_l^m$ denotes the value of the dependent
variable $u$ at the point $(l,m)\in \mathbb{Z}^2$. Attention is
restricted to difference equations on the quad-graph, which are
equations of the form
\[
F(l,m,u_l^m,u_{l+1}^m,u_l^{m+1},u_{l+1}^{m+1})=0.
\]
We also assume that this expression can be solved for any one of
the values of $u$; in particular, we write
\begin{equation}
u_{l+1}^{m+1} = \omega(l,m,u_l^m,u_{l+1}^m,u_l^{m+1}). \label{2.3}
\end{equation}
Such a nonlinear partial difference equation is said to be completely integrable if it
arises from the  compatibility condition of a system of linear difference
equations:
\begin{equation}
\left[
\begin{array}{c}
v_{l+1}^m(k) \\
w_{l+1}^m(k)
\end{array}
 \right] = L(l,m;k)  \left[
\begin{array}{c}
v_{l}^m(k) \\
w_{l}^m(k)
\end{array}
 \right], \ \ \ \   \left[
\begin{array}{c}
v_{l}^{m+1}(k) \\
w_{l}^{m+1}(k)
\end{array}
 \right] = M(l,m;k)  \left[
\begin{array}{c}
v_{l}^m(k) \\
w_{l}^m(k)
\end{array}
 \right].\label{2.1}\end{equation}
Here $v_l^m$ and $w_l^m$ are wave functions defined on the
quad-graph (or at the nodes of a two-dimensional lattice) as
functions of a spectral parameter $k$. The $2\times 2$ Lax matrices
$L(l,m;k)$ and $M(l,m;k)$ describe the change in the wave functions
under a horizontal and vertical shift respectively. They depend upon
the spectral parameter $k$ and on $u$, which plays the role of a
potential. As (\ref{2.1}) gives two different ways to express
$v_{l+1}^{m+1}(k)$ and $w_{l+1}^{m+1}(k)$ in terms of $v_l^m(k)$ and
$w_l^m(k)$, this leads to the compatibility condition
\begin{equation}
M(l+1,m;k)L(l,m;k) - L(l,m+1;k)M(l,m;k) = 0. \label{2.2}
\end{equation}
We assume that $L(l,m;k)$ and $M(l,m;k)$ depend on the potential only through $(u_l^m,u_{l+1}^m)$
and $(u_l^m,u_{l}^{m+1})$ respectively, and that the compatibility condition (\ref{2.2}) yields
the integrable quad-graph equation (\ref{2.3}).

\subsection{Lax pair compatibility conditions for the nonautonomous dKdV equation}
Generalizing the Lax pair for the dKdV equation, we consider Lax matrices $L(l,m;k)$ and $M(l,m;k)$ of the form
\begin{equation}
\begin{array}{l}
L(l,m;k) =  \left[\begin{array}{cc}a_{l+1}^m-u_{l+1}^m &  1 \\ k^2 +\alpha_l^m + (a_{l+1}^m-u_{l+1}^m)(c_l^m+u_l^m)
 &\quad c_l^m+u_l^m\end{array}
 \right],\\
M(l,m;k) =  \left[\begin{array}{cc}
b_l^{m+1}-u_l^{m+1} &  1  \\
k^2 + \beta_l^m + (b_l^{m+1}-u_l^{m+1})(d_l^m+u_l^m)\quad  & d_l^m+u_l^m
\end{array}
 \right].\end{array}\label{2.24}\end{equation}
At present, we regard $\alpha,\beta,a,b,c,d$ as arbitrary functions
of $l$ and $m$ that may occur in the nonautonomous dKdV equation.
The compatibility condition (\ref{2.2}) for the Lax matrices
(\ref{2.24}) has four components that constrain the arbitrary
functions.

Components (1,1) and (1,2) of the compatibility condition give
\begin{equation}
u_{l+1}^{m+1}= u_l^m +  c_l^m + b_{l+1}^{m+1}+ \frac{\alpha_l^m-\beta_l^m}{u_l^{m+1} -
u_{l+1}^m - b_l^{m+1}+ a_{l+1}^m} \label{2.25}\end{equation}
and
\begin{equation}
a_{l+1}^{m+1} + d_l^m = c_l^m + b_{l+1}^{m+1}.\label{con1}
\end{equation}
Substituting these results into the (2,1) component of the compatibility condition yields
\begin{gather}
\beta_{l+1}^m=\beta_l^m,~~~\alpha_l^{m+1} = \alpha_l^m, \label{2.26}\\
b_l^{m+1}+c_l^{m+1} = d_{l+1}^m + a_{l+1}^m. \label{2.28}
\end{gather}
Finally the (2,2) component produces no additional constraints. Equations (\ref{2.26}) show that
\begin{equation}
\alpha_l^m = f(l),~~~\beta_l^m = g(m),\label{2.29}\end{equation}
for arbitrary functions $f$ and $g$.
To proceed further we need to solve (\ref{con1}) and (\ref{2.28}). To do this we define new variables
$\phi_l^m$ and $\psi_l^m$ by
\begin{align}
\phi_l^m=a_l^m-b_l^m,~~~\phi_{l+1}^{m+1} = c_l^m-d_l^m,\\
\psi_{l+1}^m=b_l^m+c_l^m,~~~\psi_l^{m+1}=d_l^m + a_l^m.
\end{align}
Then (\ref{con1}) and (\ref{2.28}) imply that
\[
\psi_l^{m+1}-\psi^m_{l+1}=\phi_l^m-\phi_{l+1}^{m+1}.
\]
Let $\lambda_l^m$ be a solution of
\[
\lambda_l^m-\lambda_{l-1}^{m+1}=\phi_l^m;
\]
then
\begin{equation}
\psi_l^{m+1}-\psi_{l+1}^m=\lambda_l^m-\lambda_{l-1}^{m+1}-\lambda_{l+1}^{m+1}+\lambda_{l}^{m+2}.\label{con2}
\end{equation}
Note that $\lambda_l^m$ is defined up to an arbitrary function of $l+m$. The solution of (\ref{con2}) is
\[
\psi_l^m=\lambda_l^{m+1}-\lambda_{l-1}^m+h(l+m),
\]
where $h(l+m)$ is an arbitrary function. We can eliminate $h(l+m)$ by replacing $\lambda_l^m$ by $\lambda_l^m+H(l+m)$, where $H(l+m)$ is a solution of
\[
H(l+m+1)-H(l+m-1)+h(l+m)=0.
\]
Finally (\ref{2.25}) can be rewritten as
\begin{multline}
(u_{l+1}^{m+1}-u_l^m-b_{l+1}^{m+1}+b_l^m-\lambda_{l+1}^{m+1}+\lambda_l^m)(u_l^{m+1}-u_{l+1}^m
-b_l^{m+1}+b_{l+1}^m+\lambda_{l+1}^m-\lambda_l^{m+1})\\
=f(l)-g(m). \label{con3}
\end{multline}
The transformation
\[
u_l^m\mapsto u_l^m+b_l^m+\lambda_l^m
\]
maps (\ref{con3}) into the standard dKdV equation
\begin{equation}
(u_{l+1}^{m+1}-u_l^m)(u_{l+1}^m-u_l^{m+1})=f(l)-g(m).\label{dkdv10}
\end{equation}
Thus we find that all nonautonomous dKdV equations which admit the
Lax pair representation (\ref{2.24}) can be mapped into
(\ref{dkdv10}) by a point transformation.

\subsection{Lax pair compatibility conditions for the nonautonomous dmKdV equation}
By analogy with the autonomous dmKdV equation, let
\begin{equation}
\begin{array}{l}
\displaystyle L(l,m;k) =  \left[
\begin{array}{cc}
\displaystyle p_l^m &\quad
\displaystyle-a_l^m\ u_{l + 1}^m \\
\displaystyle-\frac {k^2 \ b_l^m}{u_l^m}  &
\displaystyle\frac {r_l^m \ u_{l + 1}^m}{u_l^m}
\end{array}
 \right],\\ \ M(l,m;k) =  \left[
\begin{array}{cc}
\displaystyle
q_l^m &
\displaystyle-d_l^m \ u_l^{m + 1}\\
\displaystyle- \frac {k^2c_l^m}{u_l^m}  &\quad
\displaystyle\frac {s_l^m\ u_l^{m + 1}}{u_l^m}
\end{array}
 \right].\end{array}\label{2.4}\end{equation}
Here $a,b,c,d,p,q,r $ and $s$ are
arbitrary functions of $l$ and $m$ that may occur in the dmKdV equation.
The (2,1) and (1,2) components of the compatibility condition (\ref{2.2}) give
\begin{align}
u_{l+1}^{m+1}= u_l^m\frac{[b_l^{m+1} \ q_l^m u_{l+1}^m - p_l^m c_{l+1}^m
\ u_l^{m+1}]}{[b_l^m \ s_{l+1}^m\ u_l^{m+1} - r_l^{m+1} c_l^m \
u_{l+1}^m]}\,,\label{2.5}\\
u_{l+1}^{m+1}=u_l^m \frac{[a_l^m q_{l+1}^m \ u_{l+1}^m
-p_l^{m+1}d_l^mu_l^{m+1}]} {[a_l^{m+1}\ s_l^m \ u_l^{m+1}-
r_l^md_{l+1}^m\ u_{l+1}^m]}\,. \label{2.6}\end{align}

Then the (1,1) and (2,2) components of the compatibility
condition yield the constraints
\begin{align}
\frac {q_{l+1}^m}{q_l^m}&=\frac{p_l^{m+1}}{p_l^m}\,,&\frac{s_{l+1}^m}{s_l^m}&=\frac {r_l^{m+1}}{r_l^m}\,,&\label{2.7}\\
\frac{b_l^{m+1}}{c_{l+1}^m}&=\frac {a_l^m}{d_l^m}\,,& \frac {d_{l+1}^m}{ a_l^{m+1}}&=\frac {c_l^m}{b_l^m}\,.& \label{2.75}
\end{align}
Combining (\ref{2.5}), (\ref{2.6}) and (\ref{2.7}) gives one further constraint on the unknown functions:
\begin{equation}
\frac{b_l^m\ a_l^m}{p_l^m \ r_l^m} = \frac{b_l^{m+1}\
a_l^{m+1}}{p_l^{m+1}\ r_l^{m+1}}\,. \label{2.8}\end{equation}

In Appendix A, we solve the constraints (\ref{2.7}), (\ref{2.75}) and (\ref{2.8}), obtaining a nonautonomous version of the dmKdV equation. This can be mapped by a point transformation into the standard dmKdV equation:
\[
u_{l+1}^{m+1}= u_l^m \frac{[g(m)u_{l+1}^m -f(l)u_l^{m+1}]}{[g(m)u_l^{m+1}-f(l)u_{l+1}^m]}\,.
\]
Here $f(l)$ and $g(m)$ are arbitrary functions.

\section{The ultra-local singularity confinement criterion}
A difference equation is said to passes the singularity confinement
criterion (which can be viewed as a discrete version of the
Painlev\'{e} Property) if every singularity is confined to a finite
number of iterations \cite{8,11,14}. Suppose that, in solving an
initial-value problem for (\ref{2.3}), a singularity occurs at the
point $(l,m)$, that is
\begin{equation}
u_l^m = O(\epsilon)\qquad \text{or}\qquad u_l^m = O \left(\frac{1}{\epsilon}\right)\qquad \text{as}\
 \epsilon \rightarrow 0.
 \label{3.1}\end{equation}
The ultra-local singularity confinement criterion requires that
\[\begin{array}{l}
u_{l+1}^{m+1} = O(1),\qquad u_{l+1}^{m-1} = O(1),\\
u_{l-1}^{m+1} = O(1),\qquad u_{l-1}^{m-1} = O(1).
\end{array}
\]
We can investigate the consequences of this criterion by solving the quad-graph equation (\ref{2.3}) for each of the
variables $u_i^j$ in turn and applying the shift operators appropriately.

\subsection{Ultra-local singularity confinement conditions for a nonautonomous dKdV equation}
Consider a nonautonomous version of the dKdV equation that is of the form
\begin{equation}
(u_{l+1}^{m+1} - u_l^m + B_l^m)(u_{l+1}^m-u_l^{m+1}+A_l^m) = C_l^m,\label{3.20}\end{equation}
where $A_l^m,B_l^m$ and $C_l^m$ are arbitrary functions. This is consistent with the
generalization (\ref{2.25}) that arises from the Lax pair.
Singularity confinement with $u_l^m = O(\epsilon)$ gives no constraints on these functions.
However, if $u_l^m = O(1/\epsilon)$, nontrivial conditions are obtained, as follows. Equation
(\ref{3.20}) can be written as
\begin{align*}
u_{l+1}^{m+1}&=u_l^m-B_l^m+\frac{C_l^m}{u_{l+1}^m-u_l^{m+1}+A_l^m}\,,\\
u_{l-1}^{m-1}&=u_l^m+B_{l-1}^{m-1}-\frac{C_{l-1}^{m-1}}{u_l^{m-1}-u_{l-1}^m+A_{l-1}^{m-1}}\,,\\
u_{l+1}^{m-1}&=u_l^m-A_l^{m-1}+\frac{C_l^{m-1}}{u_{l+1}^m-u_l^{m-1}+B_l^{m-1}}\,,\\
u_{l-1}^{m+1}&=u_l^m+A_{l-1}^m-\frac{C_{l-1}^m}{u_l^{m+1}-u_{l-1}^m+B_{l-1}^m}\,.
\end{align*}
Therefore
\begin{align*}
u_{l+1}^{m+1}=O(1)&~\Rightarrow~u_{l+1}^m-u_l^{m+1}=-A_l^m+O(\epsilon),\\
u_{l-1}^{m-1}=O(1)&~\Rightarrow~u_l^{m-1}-u_{l-1}^m=-A_{l-1}^{m-1}+O(\epsilon),\\
u_{l+1}^{m-1}=O(1)&~\Rightarrow~u_{l+1}^m-u_l^{m-1}=-B_l^{m-1}+O(\epsilon),\\
u_{l-1}^{m+1}=O(1)&~\Rightarrow~u_l^{m+1}-u_{l-1}^m=-B_{l-1}^m+O(\epsilon).
\end{align*}
From these conditions we obtain
\begin{equation}
A_{l-1}^{m-1}-A_l^m =  B_{l-1}^m - B_l^{m-1}.\label{3.25}\end{equation}
Equation (\ref{3.20}) can also be rewritten as
\begin{align}
u_l^{m+1}=&u_{l-1}^m-B_{l-1}^m+\frac{C_{l-1}^m}{u_l^m-u_{l-1}^{m+1}+A_{l-1}^m}\,,\label{dkdvsc1}\\
u_l^{m-1}=&u_{l+1}^m+B_l^{m-1}-\frac{C_l^{m-1}}{u_{l+1}^{m-1}-u_l^m+A_l^{m-1}}\,,\label{dkdvsc2}\\
u_l^{m-1}=&u_{l-1}^m-A_{l-1}^{m-1}+\frac{C_{l-1}^{m-1}}{u_l^m-u_{l-1}^{m-1}+B_{l-1}^{m-1}}\,,\label{dkdvsc3}\\
u_l^{m+1}=&u_{l-1}^m+A_l^m-\frac{C_l^m}{u_{l+1}^{m+1}-u_l^m+B_l^m}\,.\label{dkdvsc4}
\end{align}
Combining (\ref{dkdvsc1})--(\ref{dkdvsc4}) with (\ref{3.25}) gives us
\begin{multline*}
\frac{C_{l-1}^{m-1}}{u_l^m-u_{l-1}^{m-1}+B_{l-1}^{m-1}}+\frac{C_l^{m-1}}{u_{l+1}^{m-1}-u_l^m+A_l^{m-1}}
-\frac{C_{l-1}^m}{u_l^m-u_{l-1}^{m+1}+A_{l-1}^m}\\-\frac{C_l^m}{u_{l+1}^{m+1}-u_l^m+B_l^m}=0.
\end{multline*}
If $u_l^m=O(1/\epsilon)$ then this yields
\begin{equation}
u_i^j=\frac{1}{\epsilon}(C_{l-1}^m+C_l^{m-1}-C_{l-1}^{m-1}-C_l^m)+O(1),
\label{lc1}\end{equation}
where $u_i^j$ is any of $\{u_{l+1}^{m+1},~u_{l-1}^{m+1},~u_{l+1}^{m-1},~u_{l-1}^{m-1}\}$. Therefore
\begin{equation}
C_{l-1}^m+C_l^{m-1}-C_{l-1}^{m-1}-C_l^m=0.\label{dkdvsc5}
\end{equation}
The general solutions of (\ref{3.25}) and (\ref{dkdvsc5}) are
\begin{align}
A_l^m&=\lambda_{l+1}^m-\lambda_l^{m+1},\\
B_l^m&=\lambda_{l+1}^{m+1}-\lambda_l^m+h(l+m),\label{lc2}\\C_l^m&=f(l)-g(m),
\end{align}
where $\lambda_l^m,~h(l+m),~f(l),~g(m)$ are arbitrary functions. After substitution of (\ref{lc2}) into (\ref{3.20})
we obtain
\begin{multline}
(u_{l+1}^{m+1} - u_l^m +\lambda_{l+1}^{m+1}- \lambda_l^m+h(l+m))(u_{l+1}^m-u_l^{m+1}+\lambda_{l+1}^m-\lambda_l^{m+1})\\
= f(l)-g(m).\label{dkdvsc6}
\end{multline}
The transformation
\[
u_l^m\mapsto u_l^m-\lambda_l^m+H(l+m),
\]
where $H(l+m)$ is the solution of
\[
H(l+m+2)-H(l+m)+h(l+m)=0,
\]
maps (\ref{dkdvsc6}) into the standard dKdV
\begin{equation}
(u_{l+1}^{m+1} - u_l^m)(u_{l+1}^m-u_l^{m+1}) = f(l)-g(m).\label{dkdv20}
\end{equation}
Thus every nonautonomous dKdV equation of the form (\ref{3.20}) that
satisfies the ultra-local singularity confinement criterion can be transformed into (\ref{dkdv20}).

\subsection{Ultra-local singularity confinement conditions for the dmKdV equation}
We consider the following nonautonomous generalization of the dmKdV equation:
\begin{equation}
u_{l+1}^{m+1}=u_l^m \frac{[A_l^m\ u_{l+1}^m -B_l^m
u_l^{m+1}]}{[u_l^{m+1}-C_l^m\ u_{l+1}^m]},\label{mdkdv10}\end{equation}
where $A_l^m,~B_l^m$ and $C_l^m$ are arbitrary functions.
This equation can be written as
\begin{align*}
u_{l+1}^{m+1}[u_l^{m+1}-C_l^mu_{l+1}^m]&=u_l^m[A_l^mu_{l+1}^m-B_l^mu_l^{m+1}],\\
u_{l-1}^{m-1}[A_{l-1}^{m-1}u_l^{m-1}-B_{l-1}^{m-1}u_{l-1}^m] &=u_l^m[u_{l-1}^m-C_{l-1}^{m-1}u_l^{m-1}],\\
u_{l+1}^{m-1}[A_l^{m-1}u_l^{m-1}+C_l^{m-1}u_{l+1}^{m}]&=u_l^m[u_{l+1}^m+B_l^{m-1}u_l^{m-1}],\\
u_{l-1}^{m+1}[u_l^{m+1}+B_{l-1}^mu_{l-1}^m]&=u_l^m[A_{l-1}^mu_{l-1}^m+C_{l-1}^mu_l^{m+1}].\end{align*}
If $u_l^m = O(\epsilon)$ then
\begin{align*}
u_{l+1}^{m+1}=O(1)&~\Rightarrow~\frac{u_{l+1}^m}{u_l^{m+1}} = \frac{1}{C_l^m} + O(\epsilon),\\
u_{l-1}^{m-1}=O(1)&~\Rightarrow~\frac{u_l^{m-1}}{u_{l-1}^m} =\frac{B_{l-1}^{m-1}}{A_{l-1}^{m-1}}+O(\epsilon),\\
u_{l+1}^{m-1}=O(1)&~\Rightarrow~\frac{u_{l+1}^m}{u_l^{m-1}} = -\frac{A_l^{m-1}}{C_l^{m-1}} + O(\epsilon),\\
u_{l-1}^{m+1}=O(1)&~\Rightarrow~\frac{u_l^{m+1}}{u_{l-1}^m} = -B_{l-1}^m+O(\epsilon).
\end{align*}
Consequently
\begin{equation}\frac{A_{l-1}^{m-1}}{A_l^{m-1}}-\frac{B_{l-1}^{m-1}\ C_l^m}{B_{l-1}^m\
C_l^{m-1}} = 0. \label{3.13}\end{equation}
If $u_l^m = O(1/\epsilon)$ then
\begin{align*}
u_{l+1}^{m+1}=O(1)&~\Rightarrow~\frac{u_{l+1}^m}{u_l^{m+1}}= \frac{B_l^m}{A_l^m} +O(\epsilon),\\
u_{l-1}^{m-1}=O(1)&~\Rightarrow~\frac{u_{l-1}^m}{u_l^{m-1}} = C_{l-1}^{m-1} + O(\epsilon),\\
u_{l+1}^{m-1}=O(1)&~\Rightarrow~\frac{u_{l+1}^m}{u_l^{m-1}} = -B_l^{m-1} + O(\epsilon), \\
u_{l-1}^{m+1}=O(1)&~\Rightarrow~\frac{u_l^{m+1}}{u_{l-1}^m} = - \frac{A_{l-1}^m}{C_{l-1}^m} +O(\epsilon).
\end{align*}
Therefore
\begin{equation}
\frac{A_{l-1}^m}{A_l^m}-\frac{B_l^{m-1}\ C_{l-1}^m}{B_l^m\ C_{l-1}^{m-1}}= 0. \label{3.12}\end{equation}

Thus the nonautonomous dmKdV
meets the ultra-local singularity confinement criterion only if
(\ref{3.13}) and (\ref{3.12}) are satisfied.
By solving these equations we obtain a nonautonomous version of the dmKdV equation that satisfies the
ultra-local singularity confinement conditions (for details see Appendix B). This equation can be
transformed to the standard dmKdV equation,
\[
u_{l+1}^{m+1}= u_l^m \frac{[g(m)u_{l+1}^m
-f(l)u_l^{m+1}]}{[g(m)u_l^{m+1}-f(l)u_{l+1}^m]}\,
\]
where $f(l)$ and $g(m)$ are arbitrary functions.

\section{Conservation laws}
In this section, we find conservation laws for quad-graph equations
by the direct method, which is explained fully in \cite{10,RH0}.
Three-point conservation laws have components
\begin{align*}
F&=F(l,m,u_l^m,u_l^{m+1}),\\
G&=G(l,m,u_l^m,u_{l+1}^m),
\end{align*}
that satisfy the following functional equation on solutions of the given quad-graph equation:
\begin{multline}
F(l+1,m,u_{l+1}^m,u_{l+1}^{m+1})-F(l,m,u_l^m,u_l^{m+1})+G(l,m+1,u_l^{m+1},u_{l+1}^{m+1})\\-G(l,m,u_l^m,u_{l+1}^m)=0.
\label{claw}
\end{multline}
To find these conservation laws, we first substitute
\[u_{l+1}^{m+1}=\omega(l,m,u_l^m,u_{l+1}^m,u_l^{m+1})\]
into (\ref{claw}). The resulting equation involves $u_l^m,u_{l+1}^m$ and $u_l^{m+1}$, but each instance of $F$ and $G$
depends on only two continuous arguments. Therefore we can eliminate terms by repeated differentiation until a partial
differential equation for $F$ is obtained. Having solved that, it is simple to work up the hierarchy of
functional-differential consequences of (\ref{claw}) until the general solution has been obtained. The same process
can also be used to find higher conservation laws. However, the existence of three-point conservation
laws is sufficient to classify the nonautonomous dKdV and dmKdV equations.

The calculations are immensely complicated, and computer algebra is essential; we have used the computer algebra
system MAPLE. For brevity, we sketch the results of the computations, without giving full details.

\subsection{Conservation laws for the nonautonomous dKdV}
The generalization of the nonautonomous dKdV equation (\ref{3.20}) can be transformed to
\begin{equation}
(u_{l+1}^{m+1}- u_l^m + \bar{B}_l^m)(u_l^{m+1} - u_{l+1}^m)=C_l^m,\label{sdkdv1}
\end{equation}
by the point transformation
\[
u_l^m\mapsto u_l^m+Q_l^m,
\]
where $Q_l^m$ is a solution of
\[
Q_{l+1}^m-Q_l^{m+1}+A_l^m=0,
\]
and where
\[\bar{B}_l^m=B_l^m+Q_{l+1}^{m+1}-Q_l^m.\]
This simplification greatly speeds up the computations, without affecting the number of independent
conservation laws that exist. When the direct method is used to reduce (\ref{sdkdv1}) to a partial
differential equation, we find that
\begin{footnotesize}
\begin{align*}
F=&8\,(u_l^m)^{2}u_l^{m+1}\zeta_l^m\\
+&4\,u_l^{m+1} \left( u_l^{m+1}(u_l^m)^{2}\nu_l^m+2\,u_l^{m+1}u_l^m\xi_l^m+4
\,\mu_l^mu_l^m-2\,\zeta_l^{m+1}C_l^m+2\,C_l^{m+1}\zeta
_l^{m+1} \right)\\&+2\,{\frac {u_l^{m+1}T
 \left(\nu_l^mC_l^{m+1}-\nu_l^mC_l^m+2\,\zeta_l^{m+1}u_l^{m+1}+2\,\mu_l^{m+1}-2\,\mu_l^m \right) }
 {\zeta_l^m}}+{\frac {\nu_l^m(u_l^{m+1})^{2} T ^{2}}{{\zeta_l^m}^{2}}}\,,\\
G=&-8\,u_l^m \left( u_{l+1}^mu_l^m-C_l^m \right) \zeta_l^m\\
-&4\,u_{l+1}^m \left( 2\,\zeta_l^{m+1}C_l^m+4\,\mu_l^mu_l^m+u_{l+1}^m\nu_l^m(u_l^m)^{2}-2\,
\nu_l^mC_l^mu_l^m+2\,u_{l+1}^m\xi_l^mu_l^m \right)\\&-2\,{\frac {u_{l+1}^m T  \left( \nu_l^mC_l^m+\nu_l^mC
_l^{m+1}-2\,\mu_l^m+2\,u_{l+1}^m\zeta_l^{m+1}+2\,\mu_l^{m+1} \right) }
{\zeta_l^m}}-{\frac {(u_{l+1}^m)^{2}\nu_l^m
 T ^{2}}{(\zeta_l^m)^{2}}}\,,
\end{align*}
\end{footnotesize}
\!\!\!where $T=\nu_l^mC_l^m-\nu_l^mC_l^{m+1}-2\,\mu_l^m-2\,\mu_l^{m+1}$.
Here $\xi_l^m,\mu_l^n,\nu_l^m$ and $\zeta_l^m$ are functions which satisfy the following constraints:
\begin{align}
&\xi_{l+1}^m=-\zeta_l^m,~~\zeta_{l+1}^m=\zeta_l^{m+1},~~\nu_{l+1}^m=-\nu_l^m,~~\nu_l^{m+1}=-\nu_l^m,\label{CLxi}\\
&2\xi_l^m\zeta_l^m+2\zeta_l^{m+1}\zeta_l^m=-4\nu_l^m\mu_l^m-(\nu_l^m)^2C_l^m,\label{CLmu}\\
&\bar{B}_l^m=-\frac{4\mu_l^m+\nu_l^mC_l^m}{2\zeta_l^m}\,.\label{CLnu}
\end{align}
Note: at this stage, we have not completed the direct method calculation of the conservation laws, but the above
necessary conditions lead to a substantial further simplification of the problem.

The general solution of the system (\ref{CLxi}) is
\begin{equation}
\zeta_l^m=H(l+m),~~~\xi_l^m=-H(l+m-1),~~~\nu_l^m=c_1(-1)^{l+m},\label{CLzeta}
\end{equation}
where $H(l+m)$ is an arbitrary nonzero function and $c_1$ is an arbitrary nonzero constant. Combining these
results with (\ref{CLmu}), and (\ref{CLnu}), we obtain
\[
\bar{B}_l^m=\frac{2}{c_1}(-1)^{l+m}(H(l+m+1)-H(l+m-1)).
\]
Therefore three-point conservation laws exist only if the nonautonomous dKdV equation is of the form
\begin{equation}
\left(u_{l+1}^{m+1}- u_l^m + \frac{2}{c_1}(-1)^{l+m}[H(l+m+1)-H(l+m-1)]\right)(u_l^{m+1} - u_{l+1}^m)=C_l^m.
\label{sdkdv2}
\end{equation}
This equation is mapped by the point transformation
\[
u_l^m\mapsto u_l^m-\frac{2}{c_1}(-1)^{l+m}H(l+m-1)
\]
to
\begin{equation}
(u_{l+1}^{m+1}- u_l^m)(u_l^{m+1} - u_{l+1}^m)=C_l^m.\label{sdkdv3}
\end{equation}
Therefore it is enough to seek conservation laws of (\ref{sdkdv3}). Applying the full direct
method to (\ref{sdkdv3}) gives us one further condition on $C_l^m$:
\[
C_{l+1}^{m+1}-C_{l+1}^m-C_l^{m+1}+C_l^m=0.
\]
Consequently all nonautonomous dKdV equations that have nontrivial conservation laws can be mapped
to
\begin{equation}
(u_{l+1}^{m+1}-u_l^m)(u_{l+1}^m-u_l^{m+1})=f(l)-g(m),\label{sdkdv4}
\end{equation}
whose three-point conservation laws are are
\begin{enumerate}
\item $F=(-1) ^{l+m} \left( 2u_l^mu_l^{m+1}+g(m) \right)$,\\
$G=-(-1) ^{l+m} \left( 2u_l^mu_{l+1}^m+f(l) \right)$,
\item $F=\left( u_l^m-u_l^{m+1} \right) \left( u_l^mu_l^{m+1}+g(m) \right)$,\\
$G=- \left( u_l^m-u_{l+1}^m \right) \left( u_l^mu_{l+1}^m+f(l) \right)$,
\item $F=(-1) ^{l+m} \left( u_l^m+u_l^{m+1} \right)\left( u_l^mu_l^{m+1}+g(m) \right)$,\\
$G=-(-1)^{l+m} \left( u_l^m+u_{l+1}^m \right) \left(u_l^mu_{l+1}^m+f(l) \right)$
\item $F=(-1)^{l+m}\left(2(u_l^mu_l^{m+1})^2+4g(m)u_l^mu_l^{m+1}+(g(m))^2\right)$,\\
$G=-(-1)^{l+m}\left( 2(u_l^mu_{l+1}^m)^2+4f(l)u_l^mu_{l+1}^m+(f(l))^2 \right)$.
\end{enumerate}

\subsection{Conservation laws for the nonautonomous dmKdV equation}
In the same way (but with fewer details), we apply the direct method to the nonautonomous dmKdV equation
(\ref{mdkdv10}). Then the components $F$ and $G$ are of the form
\begin{align*}
F&=\nu_l^mu_l^mu_l^{m+1}-\xi_l^m\frac{u_l^{m+1}}{u_l^m}-\zeta\frac{u_l^m}{u_l^{m+1}}
+\frac{\mu}{u_l^mu_l^{m+1}}\,,\\
G&=-\nu_l^m\frac{A_l^m}{B_l^m}u_l^mu_{l+1}^m+\xi_l^mC_l^m\frac{u_{l+1}^m}{u_l^m}
+\zeta_l^m\frac{B_l^m}{A_l^m}\frac{u_l^m}{u_{l+1}^m}-\frac{\mu_l^m}{C_l^mu_l^mu_{l+1}^m}\,,
\end{align*}
where the functions $\xi,\mu,\nu$ and $\zeta$ satisfy the constraints
\begin{align*}
\xi_{l+1}^m\zeta_{l+1}^m&=\xi_l^m\zeta_l^m,\\
\mu_{l+1}^m\nu_{l+1}^m&=\mu_l^m\nu_l^m,\\
{C_l^{m+1}}/{C_l^m}&={\xi_{l+1}^m}/{\xi_l^{m+1}},\\
A_l^m&={\zeta_{l+1}^m}/{\xi_l^m},\label{cond4}\\
\xi_l^{m+1}\mu_l^{m+1}&=\mu_l^m\zeta_l^m,\\
\nu_l^{m+1}\zeta_l^{m+1}&=\xi_l^m\nu_l^m,\\
\nu_{l+1}^mB_l^m&=\nu_l^mC_l^m.
\end{align*}
The general solution of these constraints is very messy. However, it yields the result that the
only nonautonomous dmKdV equations with nonzero $A,~B,~C$ that admit conservation laws can be transformed
to the standard dmKdV equation
\[
u_{l+1}^{m+1}= u_l^m \frac{[g(m)u_{l+1}^m -f(l)u_l^{m+1}]}{[g(m)u_l^{m+1}-f(l)u_{l+1}^m]}.
\]
The three-point conservation laws for this equation are
\begin{enumerate}
\item $\displaystyle F=\frac{u_l^mu_l^{m+1}}{g(m)}$\,,\\
$\displaystyle G=-\frac{u_{l+1}^mu_l^m}{f(l)}\,,$
\item $\displaystyle F={\frac{1}{g(m)u_l^mu_l^{m+1}}}$\,,\\
$\displaystyle G=-{\frac{1}{f(l)u_l^mu_{l+1}^m}}\,$,
\item $\displaystyle F=g(m)\left(\frac{u_l^m}{u_l^{m+1}}+\frac{u_l^{m+1}}{u_l^m}  \right)$,\\
$\displaystyle G=-f(l)\left(\frac{u_l^m}{u_{l+1}^m}+\frac{u_{l+1}^m}{u_l^m}  \right)$,
\item $\displaystyle F=(-1)^{l+m}g(m)\left(\frac{u_l^m}{u_l^{m+1}}-\frac{u_l^{m+1}}{u_l^m}  \right)$,\\
$\displaystyle G=-(-1)^{l+m}f(l)\left(\frac{u_l^m}{u_{l+1}^m}-\frac{u_{l+1}^m}{u_l^m}  \right)$.
\end{enumerate}
\section{Conclusion}
We have studied the existence of nonautonomous versions of dKdV and dmKdV that satisfy the Lax pair
constraints, the ultra-local singularity confinement criterion and the conditions under which
a quad-graph admits three-point conservation laws. Each of these conditions imply that all integrable
systems of these forms are related to the standard dKdV equation
\begin{equation*}
(u_{l+1}^{m+1}-u_l^m)(u_{l+1}^m-u_l^{m+1})=f(l)-g(m),
\end{equation*}
or dmKdV equation
\[
u_{l+1}^{m+1}= u_l^m \frac{[g(m)u_{l+1}^m -f(l)u_l^{m+1}]}{[g(m)u_l^{m+1}-f(l)u_{l+1}^m]}\,,
\]
respectively, by a transformation \[u_l^m\mapsto
P(l,m)u_l^m+Q(l,m),\] where $P(l,m)$ and $Q(l,m)$ are arbitrary
functions. This result is in agreement with results from the paper
 \cite{9} which uses the consistency approach. It suggests that there is a
deep connection between the various criteria for integrability of
quad-graphs. In order to find this connection more research has to
be done.

The methods that we have used in this paper can be applied to the other integrable quad-graph equations in the classification by Adler, Bobenko \& Suris \cite{9}. These equations all have the tetrahedron property; at present, it is not yet known whether there exist integrable quad-graphs without this property that are not linearizable (see \cite{Hie0,RJGT}). If such quad-graphs exist, they could also be tested by the methods in this paper.

\eject
{\bf APPENDICES}\\
\begin{appendix}
\section{Lax pair compatibility conditions for the nonautonomous dmKdV equation}
By combining equations (\ref{2.75}), we obtain
\[
\frac{a_l^{m+1}b_l^{m+1}}{a_l^mb_l^m}=\frac{c_{l+1}^md_{l+1}^m}{c_l^md_l^m}\,.
\]
This can be integrated (assuming that the domain has trivial difference cohomology) to give
\begin{align}
a_l^mb_l^m&=\frac{\psi_{l+1}^m}{\psi_l^m}\,,\label{mdkdvc1}\\
c_l^md_l^m&=\frac{\psi_l^{m+1}}{\psi_l^m}\,,\label{mdkdvc1.5}
\end{align}
where $\psi_l^m$ is an arbitrary function. Similarly from (\ref{2.7}) we have
\begin{align}
r_l^m&=\frac{\chi_{l+1}^m}{\chi_l^m}\,,& p_l^m&=\frac{\phi_{l+1}^m}{\phi_l^m}\,,&\label{mdkdvc2}\\
s_l^m&=\frac{\chi_l^{m+1}}{\chi_l^m}\,,& q_l^m&=\frac{\phi_l^{m+1}}{\phi_l^m}\,,&
\end{align}
where $\phi_l^m$ and $\chi_l^m$ are arbitrary functions.
Equation (\ref{2.8}) also can be integrated:
\begin{equation}
\frac{a_l^mb_l^m}{p_l^mr_l^m}=H(l),\label{mdkdvc5}
\end{equation}
where $H(l)$ is an arbitrary function.
By combining (\ref{mdkdvc1}), (\ref{mdkdvc2}) and (\ref{mdkdvc5}) we obtain
\begin{equation}
\frac{\psi_{l+1}^m\phi_l^m\chi_l^m}{\psi_l^m\phi_{l+1}^m\chi_{l+1}^m}=H(l).
\end{equation}
Let $H(l)=\frac{F(l+1)}{F(l)}$, where $F(l)$ is defined up to an arbitrary nonzero constant factor. Then
\begin{equation}
\frac{\phi_{l+1}^m\chi_{l+1}^mF(l+1)}{\psi_{l+1}^m}
=\frac{\phi_l^m\chi_l^mF(l)}{\psi_l^m}\,,\quad \text{and so}\ \psi_l^m=\phi_l^m\chi_l^mF(l)G(m).
\end{equation}
Therefore, from (\ref{mdkdvc1}),
\begin{equation}
a_l^m=\frac{\psi_{l+1}^m}{b_l^m\psi_l^m}=\frac{\phi_{l+1}^m\chi_{l+1}^mF(l+1)}{b_l^m\phi_l^m\chi_l^mF(l)}\,,\label{mdkdvc6}
\end{equation}
and, from (\ref{mdkdvc1.5}),
\begin{equation}
d_l^m=\frac{\psi_l^{m+1}}{c_l^m\psi_l^m}=\frac{\phi_l^{m+1}\chi_l^{m+1}G(m+1)}{c_l^m\phi_l^m\chi_l^mG(m)}\,.\label{mdkdvc7}
\end{equation}
Substituting (\ref{mdkdvc6}) and (\ref{mdkdvc7}) into the first of the remaining constraints (\ref{2.75}) gives
\begin{multline*}
[c_{l+1}^m\phi_{l+1}^m\chi_{l+1}^mF(l+1)G(m)][c_l^m\phi_l^m\chi_l^mF(l)G(m)]\\
=[b_l^{m+1}\phi_l^{m+1}\chi_l^{m+1}F(l)G(m+1)][b_l^m\phi_l^m\chi_l^mF(l)G(m)].
\end{multline*}
Solving this equation, we obtain
\begin{align}
b_l^m&=\frac{\eta_l^m\eta_{l+1}^m}{\phi_l^m\chi_l^mF(l)G(m)}\,,\quad&c_l^m&
=\frac{\eta_l^m\eta_l^{m+1}}{\phi_l^m\chi_l^mF(l)G(m)}\,,&\\
a_l^m&=\frac{\phi_{l+1}^m\chi_{l+1}^mF(l+1)G(m)}{\eta_l^m\eta_{l+1}^m}\,,\quad&d_l^m&
=\frac{\phi_l^{m+1}\chi_l^{m+1}F(l)G(m+1)}{\eta_l^m\eta_l^{m+1}}\,,&
\end{align}
where $\eta_l^m$ is an arbitrary function.
Therefore (\ref{2.5}) amounts to
\begin{equation}
u_{l+1}^{m+1}=\frac{\eta_{l+1}^{m+1}\chi_l^mu_l^m}{\eta_l^m\chi_{l+1}^{m+1}}
\left(\frac{\frac{\chi_{l+1}^m(g(m))^2}{\eta_{l+1}^m}u_{l+1}^m
-\frac{\chi_l^{m+1}(f(l))^2}{\eta_l^{m+1}}u_l^{m+1}}{\frac{\chi_l^{m+1}}{\eta_l^{m+1}}u_l^{m+1}
-\frac{\chi_{l+1}^m}{\eta_{l+1}^m}u_{l+1}^m}\right),\label{mdkdvc8}
\end{equation}
where $(f(l))^2=\frac{F(l)}{F(l+1)}$ and $(g(m))^2=\frac{G(m)}{G(m+1)}$. The transformation \[u_l^m\mapsto u_l^m\delta(m)\gamma(l)\frac{\eta_l^m}{\chi_l^m}\,,\]
where $\frac{\delta(m+1)}{\delta(m)}=g(m)$ and $\frac{\gamma(l+1)}{\gamma(l)}=f(l)$, reduces (\ref{mdkdvc8}) to the standard dmKdV equation
\[
u_{l+1}^{m+1}= u_l^m \frac{[g(m)u_{l+1}^m -f(l)u_l^{m+1}]}{[g(m)u_l^{m+1}-f(l)u_{l+1}^m]}\,.
\]
\section{Ultra-local singularity confinement conditions for the nonautonomous dmKdV equation}
The conditions (\ref{3.13}) and (\ref{3.12}) amount to
\begin{equation}
 \frac{A_{l+1}^m}{A_l^m}=\frac{B_{l+1}^mC_l^{m-1}}{B_{l+1}^{m-1}C_l^m}
 =\frac{B_l^{m+1}C_{l+1}^m}{B_l^mC_{l+1}^{m+1}}\,.\label{ulcc1}
\end{equation}
Consequently
\[
\frac{B_{l+1}^{m-1}C_{l+1}^m}{B_l^mC_l^{m-1}}=H(l),
\]
where $H(l)$ is an arbitrary function. Integrating this gives
\[
B_l^m=(f(l))^2\frac{\psi_l^{m-1}}{\psi_{l+1}^m},~~~C_l^m
=\frac{\psi_{l+1}^{m-1}}{\psi_l^m},~~\text{where}~~\left(\frac{f(l+1)}{f(l)}\right)^2=H(l).
\]
Therefore (\ref{ulcc1}) yields
\[
\frac{A_{l+1}^m}{A_l^m}=\frac{\psi_{l+2}^{m-1}\psi_l^m}{\psi_{l+2}^m\psi_l^{m-1}}\,,
\]
and so
\[ A_l^m=(g(m))^2\frac{\psi_{l+1}^{m-1}\psi_l^{m-1}}{\psi_{l+1}^m\psi_l^m}\,,\]
where $g(m)$ is an arbitrary nonzero function. Then (\ref{mdkdv10}) amounts to
\[
u_{l+1}^{m+1}=\frac{\psi_l^{m-1}u_l^m}{\psi_{l+1}^m}
\left(\frac{(g(m))^2\psi_{l+1}^{m-1}u_{l+1}^m-(f(l))^2\psi_l^mu_l^{m+1}}{\psi_l^mu_l^{m+1}
-\psi_{l+1}^{m-1}u_{l+1}^m}\right).
\]
The transformation \[u_l^m\mapsto u_l^m\frac{\delta(m)\gamma(l)}{\psi_l^{m-1}}\,
,\]
where $\frac{\delta(m+1)}{\delta(m)}=g(m)$ and $\frac{\gamma(l+1)}{\gamma(l)}=f(l)$, reduces (\ref{mdkdv10}) to the standard dmKdV equation
\[
u_{l+1}^{m+1}= u_l^m \frac{[g(m)u_{l+1}^m -f(l)u_l^{m+1}]}{[g(m)u_l^{m+1}-f(l)u_{l+1}^m]}\,.
\]

\end{appendix}

\end{document}